\documentstyle[11pt,aaspp4,flushrt]{article}
\slugcomment{\it Submitted to the Astrophysical Journal}
\lefthead{Stern et al.}
\righthead{SEXSI Gets Lucky}

\def\ser{SEXSI-SER}

\def\cf{{cf.,~}}

\def\eg{{e.g.,~}}
\def\etal{{et al.}}

\def\deg{\ifmmode {^{\circ}}\else {$^\circ$}\fi}
\def\kms{\ifmmode {\rm\,km\,s^{-1}}\else
    ${\rm\,km\,s^{-1}}$\fi}
\def\ergcm2s{\ifmmode {\rm\,ergs\,cm^{-2}\,s^{-1}}\else
    ${\rm\,ergs\,cm^{-2}\,s^{-1}}$\fi}
\def\ergAcm2s{\ifmmode {\rm\,ergs\,cm^{-2}\,s^{-1}\,\AA^{-1}}\else
    ${\rm\,ergs\,cm^{-2}\,s^{-1}\,\AA^{-1}}$\fi}
\def\ergs{\ifmmode {\rm\,ergs\,s^{-1}}\else
    ${\rm\,ergs\,s^{-1}}$\fi}
\def\kmsMpc{\ifmmode {\rm\,km\,s^{-1}\,Mpc^{-1}}\else
    ${\rm\,km\,s^{-1}\,Mpc^{-1}}$\fi}

\def\lya{Ly$\alpha$}

\def\civ{\ion{C}{4} $\lambda$1549}
\def\oii{[\ion{O}{2}] $\lambda$3727}
\def\oiii{[\ion{O}{3}] $\lambda$5007}
\def\oiiia{[\ion{O}{3}] $\lambda$4959}
\def\oiiipair{[\ion{O}{3}] $\lambda \lambda$4959,5007}
\def\nii{[\ion{N}{2}] $\lambda \lambda$6548,6584}
\def\sii{[\ion{S}{2}] $\lambda \lambda$6716,6731}

\def\spose#1{\hbox to 0pt{#1\hss}}
\def\simlt{\mathrel{\spose{\lower 3pt\hbox{$\mathchar"218$}}
     \raise 2.0pt\hbox{$\mathchar"13C$}}}
\def\simgt{\mathrel{\spose{\lower 3pt\hbox{$\mathchar"218$}}
     \raise 2.0pt\hbox{$\mathchar"13E$}}}

\def\plotfiddle#1#2#3#4#5#6#7{\centering \leavevmode
\vbox to#2{\rule{0pt}{#2}}
\includegraphics{#1}}

\begin{document}

\title{A Galaxy at $z = 6.545$ and Constraints on the Epoch of Reionization}

\author{Daniel Stern\altaffilmark{1},
Sarah A.~Yost\altaffilmark{2},
Megan E.~Eckart\altaffilmark{2},
Fiona A.~Harrison\altaffilmark{2}, \\
David J.~Helfand\altaffilmark{3},
S.G.~Djorgovski\altaffilmark{4},
Sangeeta~Malhotra\altaffilmark{5},
and James E.~Rhoads\altaffilmark{5}}

\altaffiltext{1}{Jet Propulsion Laboratory, California Institute
of Technology, Mail Stop 169-506, Pasadena, CA 91109. e-mail: {\tt
stern@zwolfkinder.jpl.nasa.gov}}

\altaffiltext{2}{Space Radiation Laboratory, California Institute of
Technology, Mail Code 220-47, Pasadena CA 91125}

\altaffiltext{3}{Columbia University, Department of Astronomy, 550 West
120th Street, New York, NY 10027}

\altaffiltext{4}{California Institute of Technology, Astronomy, Mail Code
105-24, Pasadena CA 91125}

\altaffiltext{5}{Space Telescope Science Institute, 3700 San Martin
Drive, Baltimore, MD 21218}

\begin{abstract} 

We report the discovery of a \lya-emitting galaxy at redshift $z = 6.545$
serendipitously identified in the course of spectroscopic follow-up of
hard X-ray sources on behalf of the Serendipitous Extragalactic X-Ray
Source Identification (SEXSI) survey.  The line flux of the galaxy,
$2.1 \times 10^{-17} \ergcm2s$, is similar to line fluxes probed by
narrow-band imaging surveys; the $5.2~ {\rm arcmin}^2$ surveyed implies
a surface density of $z \sim 6.5$ \lya\ emitters somewhat higher than
that inferred from narrow-band surveys.  This source marks the sixth
\lya-emitting galaxy identified at $z \sim 6.5$, a redshift putatively
beyond the epoch of reionization when the damping wings of the neutral
hydrogen of the intergalactic medium is capable of severely attenuating
\lya\ emission.  By comparing the \lya\ luminosity functions at $z \sim
5.7$ and $z \sim 6.5$, we infer that the intergalactic medium may remain
largely reionized from the local universe out to $z \sim 6.5$.

\end{abstract}


\section{Introduction}

The observation of galaxy and quasar light emitted when the Universe
was less than a billion years old has opened a window for spectroscopic
exploration of the early history of the intergalactic medium.
\markcite{Djorgovski:01b}Djorgovski {et~al.} (2001) showed a dramatic increase in the optical depth
of the \lya\ forest at $z \simgt 5.2$ and interpreted this as evidence
of the trailing edge of the cosmic reionization epoch.   Subsequent
discovery of broad, black, \lya\ absorption troughs, as predicted by
\markcite{Gunn:65}Gunn \& Peterson (1965), in the spectra of the highest redshift quasars ($z
\simgt 6$) indicate that we are beginning to probe the era when
hydrogen was reionized by an early generation of stars
\markcite{Becker:01, White:03, Fan:03}(Becker {et~al.} 2001; White {et~al.} 2003; Fan {et~al.} 2003).  Circa 2001, we thought we had
identified where the ``dark ages'' ended, when the first sources of
light in the Universe turned on with sufficient strength to dissociate
the hydrogen atoms formed when the Universe was only $\sim$300,000
years old.

More recently, our picture of the early ionization history of the
intergalactic medium (IGM) has become more complicated.  The {\it
Wilkinson Microwave Anisotropy Probe} identified a large amplitude
signal in the temperature-polarization maps of the cosmic microwave
background \markcite{Spergel:03}(Spergel {et~al.} 2003), indicating a large optical depth to
Thomson scattering.  The straightforward interpretation of this result is
that the universe became reionized at $z = 17 \pm 5$ \markcite{Kogut:03}(Kogut {et~al.} 2003).
At first glance this appears inconsistent with reionization occuring at
$z \sim 6$.  However, since a relatively small neutral fraction ($x_{\rm
HI}^{\rm IGM} \approx 0.001$) suffices to produce black Gunn-Peterson
troughs, the microwave results are not necessarily contradictory with the
quasar results:  perhaps reionization began early, and only completed
around redshift $z \sim 6$.  On the other hand, \markcite{Wyithe:04}Wyithe \& Loeb (2004a) and
\markcite{Mesinger:04}Mesinger \& Haiman (2004) note that the sizes of the \ion{H}{2} region around
the highest redshift quasars provide a stronger constraint on the IGM
neutral fraction, implying $x_{\rm HI}^{\rm IGM} > 0.1$ for typical
quasar lifetimes.

There is mounting evidence that the ionization history of the IGM is
complex; see, \eg \markcite{Loeb:01}Loeb \& Barkana (2001), \markcite{Barkana:01}Barkana \& Loeb (2001), and
\markcite{MiraldaEscude:03}Miralda-Escud\'e (2003) for recent reviews.  Several theorists have
argued that the hydrogen in the IGM could have been reionized twice
\markcite{Wyithe:03, Cen:03, Haiman:03, Somerville:03}(\eg Wyithe \& Loeb 2003; Cen 2003; Haiman \& Holder 2003; Somerville, Bullock, \&  Livio 2003).  A first
reionization occurs at $z \sim 20$, driven by the formation of massive,
zero-metallicity Population~III stars \markcite{Bromm:04}(\eg Bromm 2004).
However, their radiative and mechanical feedback may have disrupted the
subsequent star formation, possibly leading to a partial recombination
until an increasing Population~II massive star population and declining
IGM density allow the second reionization to occur at $z \sim 6$, as
indicated by the high-redshift quasar spectroscopic measurements.  How
and when reionization(s) happened, and the details of the reionization
process, have been among the most pressing questions in astrophysical
cosmology, holding many clues about the first generation of light
sources.

High-redshift \lya\ emitters offer a complementary approach to studying
the early reionization history of the IGM.  It has long been suggested
that the first sources of ultraviolet radiation, responsible for
the reionization of the Universe, should be strong \lya\ emitters
\markcite{Partridge:67}(Partridge \& Peebles 1967).  After several decades of largely unsuccessful
searches \markcite{Pritchet:94}(for a review, see Pritchet 1994), numerous \lya\ emitters
are finally being discovered at high redshifts \markcite{Dey:98,
Hu:98, Weymann:98, Rhoads:00, Ellis:01, Dawson:02, Kodaira:03, Rhoads:03,
Hu:04, Dickinson:04, Stanway:04}(\eg Dey {et~al.} 1998; Hu, Cowie, \& McMahon 1998; Weymann {et~al.} 1998; Rhoads {et~al.} 2000; Ellis {et~al.} 2001; Dawson {et~al.} 2002; Kodaira {et~al.} 2003; Rhoads {et~al.} 2003; Hu {et~al.} 2004; Dickinson {et~al.} 2004; Stanway {et~al.} 2004).  These discoveries demonstrate that
\lya-emitting galaxies do exist in significant numbers out to the current
horizon of their detectability.  The most distant sources, confirmed
out to $z = 6.578 \pm 0.002$ \markcite{Kodaira:03}(Kodaira {et~al.} 2003), are seen {\it prior}
to the quasar-derived reionization redshift of the IGM.  Since \lya\
photons injected into a neutral IGM are strongly scattered, the red
damping wing of the Gunn-Peterson trough should strongly suppress, or even
completely eliminate, detectable \lya\ emission \markcite{MiraldaEscude:98a,
MiraldaEscude:98b, Loeb:99, Barkana:04, Gnedin:04}(Miralda-Escud\'e \& Rees 1998; Miralda-Escud\'e 1998; Loeb \& Rybicki 1999; Barkana \& Loeb 2004; Gnedin \& Prada 2004).  Initially it was
thought that small \lya-emitting galaxies embedded in a neutral IGM
would be incapable of ionizing enough of their surroundings to prevent
this effect, thus implying that the detection of even a {\em single}
\lya-emitting galaxy requires that the reionization epoch occurs at a
higher redshift.  Subsequent calculations \markcite{Haiman:02}(Haiman 2002), however,
show that even for faint sources with little ionizing continuum, a
sufficiently broad ($\Delta v \simgt 300\, {\rm km}\ {\rm s}^{-1}$)
emission line can remain observable.  The transmitted fraction depends
upon the size of the local cosmological \ion{H}{2} region surrounding a
source, and therefore on the ionizing luminosity and age of the source
\markcite{Santos:04}(\eg Santos 2004) as well as on contributions from associated,
clustered sources \markcite{Wyithe:04b}(\eg Wyithe \& Loeb 2004b).  Nevertheless, we expect
a rapid decline in the observed space density of \lya\ emitters as
the reionization epoch is approached:  a statistical sample of \lya\
emitters spanning the reionization redshift should still be a useful
probe of reionization \markcite{Haiman:99, Rhoads:01c, Haiman:02}(Haiman \& Spaans 1999; Rhoads \& Malhotra 2001; Haiman 2002).  However,
such an effect may be detectable only for the fainter sources, and it
depends on many as-yet unknown details about the geometry of reionization,
intrinsic evolution of the Ly$\alpha$ luminosity function, as well as
the winds and environment of high-redshift \lya\ emitters.

Here we report the discovery of a \lya\ emitting galaxy at $z = 6.545$,
hereafter referred to as \ser, serendipitously identified in the course
of spectroscopic follow-up of hard ($2 - 10$ keV) X-ray sources from
the Serendipitous Extragalactic X-ray Source Identification program
\markcite{Harrison:03, Eckart:04}(SEXSI; Harrison {et~al.} 2003; Eckart {et~al.} 2004).  The SEXSI survey, whose
scientific goal is to understand the sources contributing to the cosmic
X-ray background, has obtained spectroscopic redshifts for $\simgt 450$
hard X-ray sources selected from 27 archival {\it Chandra X-Ray
Observatory} fields (Eckart \etal, in preparation).  By nature of this
extensive spectroscopic campaign of faint, modest surface density
sources, the SEXSI survey is also particularly well-suited to
serendipitous searches for high-redshift \lya-emitting galaxies.

\ser\ marks the sixth \lya-emitting galaxy identified at $z \sim 6.5$.
This redshift corresponds to clean window in the night sky, relatively
free of telluric emission lines, and has thus been a preferred redshift
for narrow-band imaging surveys.  Four $z \sim 6.5$ \lya\ emitters
have been confirmed from such surveys to date \markcite{Hu:02, Kodaira:03,
Rhoads:04}(Hu {et~al.} 2002; Kodaira {et~al.} 2003; Rhoads {et~al.} 2004).  A fifth $z \sim 6.5$ \lya\ emitter has recently been reported
from a slitless spectroscopy program with VLT \markcite{Kurk:04}(Kurk {et~al.} 2004), while
\markcite{Tran:04}Tran {et~al.} (2004) report on an unsuccessful narrow-band {\it spectroscopic}
search with the same telescope.

We describe our imaging and spectroscopic observations in \S 2 and discuss
the inferred physical properties of \ser\ in \S 3.1.  The implications
of this discovery with respect to the early star formation density
and the epoch of reionization comprise \S~3.2 and \S~3.3, respectively.
A coordinated paper, \markcite{Malhotra:04}Malhotra \& Rhoads (2004), provides a more detailed analysis
of the luminosity function of \lya\ emitters at $z \approx 5.7$ and $z
\approx 6.5$, as a probe of reionization epoch.  Throughout we adopt a
$\Lambda$-cosmology with $\Omega_M = 0.3$, $\Omega_\Lambda = 0.7$, and
$H_0 = 70\, \kmsMpc$.  At $z = 6.545$, such a universe is 0.82~Gyr old,
the lookback time is 93.9\% of the total age of the universe, and an
angular size of 1\farcs0 corresponds to 5.4~kpc.

\section{Observations}

\subsection{Optical Imaging}

\begin{figure}[!t]
\begin{center}
\plotfiddle{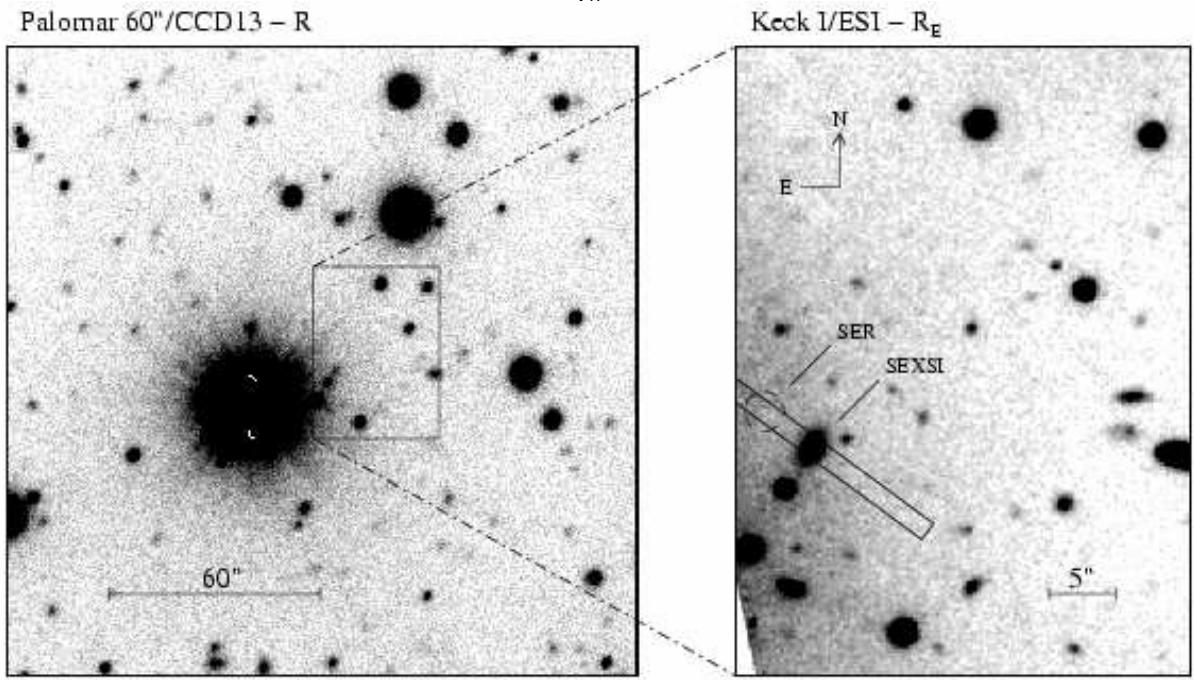}{2.9in}{0}{80}{80}{-260}{-200}
\end{center}

\caption{Optical, $R$-band images of \ser, in the MS1621.5+2640
field.  Orientation and scales are indicated.  The left-hand panel
presents shallow, wide-area $R$-band imaging from Palomar Observatory.
The right-hand panel highlights the area around \ser\ and was obtained
with the Keck~II telescope through an Ellis $R_E$-band filter.  The
Keck imaging was oriented so as to avoid the bright star ESE of \ser.
The target of the slitlet which serendipitously covered \ser\ was
CXOSEXSI~J162256.7+264103 (labeled ``SEXSI''); the orientation
(position angle of 53.7\deg) and size (1\farcs0 $\times$ 27\arcsec) of
the slitlet is indicated.  We find no optical source at or near the
location of \ser\ (labeled ``SER''), consistent with the $z = 6.5$
interpretation of the spectrum.}

\label{fig.image}
\end{figure}

\ser\ was identified in the outskirts of MS1621.5+2640
\markcite{Ellingson:97}(Ellingson {et~al.} 1997), a $z = 0.428$ galaxy cluster which was the target
of a 30~ks exposure with {\it Chandra} and is one of the 27 SEXSI
survey fields.  To support spectroscopic follow-up of X-ray sources in
the MS1621.5+2640 field, we obtained shallow, wide-field images with
the Palomar 60-inch CCD13 camera on UT 2001 May 18.  \markcite{Eckart:04}Eckart {et~al.} (2004)
presents these observations, obtained through a Kron-Cousins $R$-band
filter ($\lambda_c = 6450$~\AA; $\Delta \lambda = 1500$~\AA).  On UT
2004 April 25, we used the Echelle Spectrograph and Imager
\markcite{Sheinis:02}(ESI; Sheinis {et~al.} 2002) on the Keck~II telescope to obtain a deep,
1200~s image through an Ellis $R_E$-band filter ($\lambda_c =
6657$~\AA; $\Delta \lambda = 1200$~\AA) in photometric conditions with
0\farcs7 seeing.  Fig.~\ref{fig.image} displays these optical images.

\subsection{Keck Spectroscopy}

We identified \ser\ in spectra obtained with the Deep Imaging
Multi-Object Spectrograph \markcite{Faber:03}(DEIMOS; Faber {et~al.} 2003) on the Keck~II
telescope.  These observations, taken on UT 2003 August 24, roughly
contemporaneous with the launch of the {\it Spitzer Space Telescope},
were on the final night of a three-night run studying SEXSI sources
over multiple fields.  During this observing run we observed 19
slitmasks, each of which used 1\farcs0 wide slitlets, the 600ZD grating
($\lambda_{\rm blaze} = 7500$ \AA; $\Delta \lambda_{\rm FWHM} = 3.7$
\AA), and a GG455 order-blocking filter.  Typical exposure times were
1~hr, split into three 1200~s exposures for effective cosmic ray
removal.  We used {\tt IRAF} to process the data using standard
techniques.  Flux calibration relied upon observations of G191B2B
obtained during the same observing run.  Because of contamination of
second order light longward of 9100~\AA\ from the blue standard star,
the flux calibration at long wavelengths is extrapolated; this estimate
should be accurate to better than 10\% at 9200~\AA.

\begin{figure}[!t]
\begin{center}
\plotfiddle{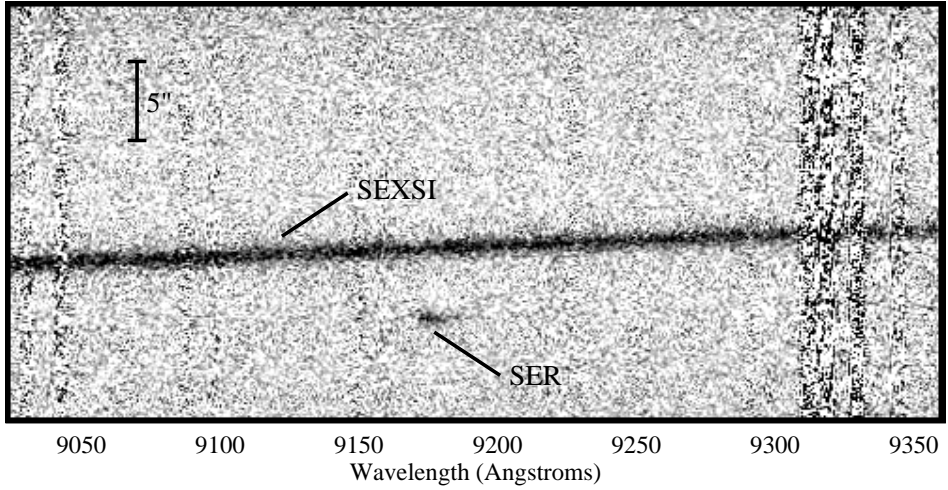}{1.9in}{0}{60}{60}{-200}{-20}
\end{center}

\caption{Two-dimensional, processed spectrum of \ser\ at $z = 6.545$,
obtained with DEIMOS on the Keck~II telescope.  The total exposure time
is 1~hr.}

\label{fig.2dspec}
\end{figure}

\begin{figure}[!t]
\begin{center}
\plotfiddle{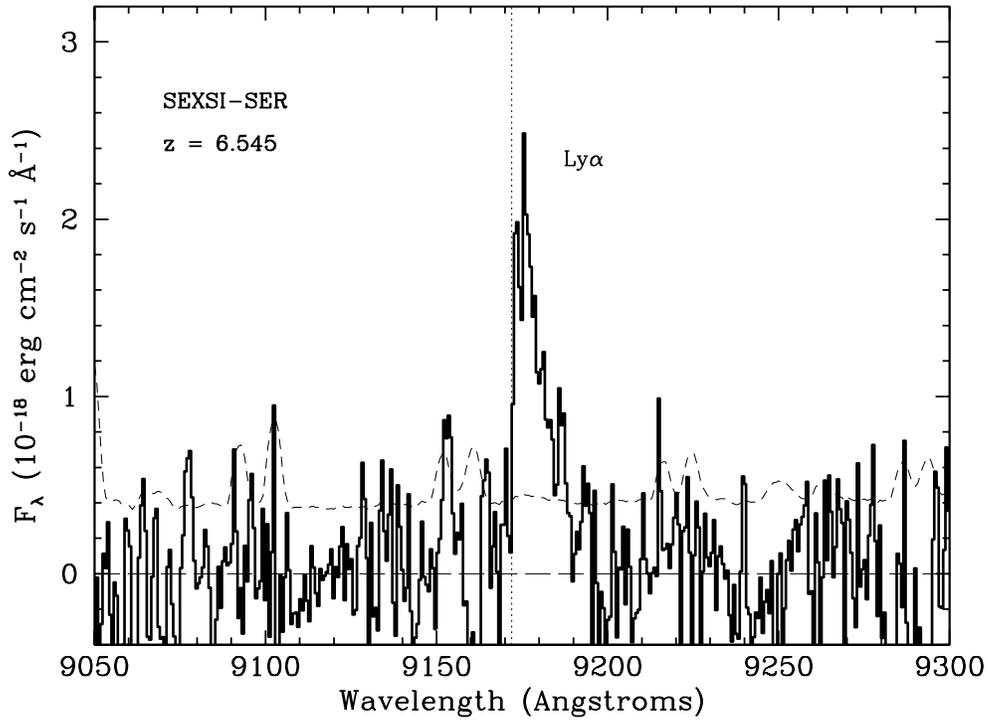}{3.4in}{-90}{50}{50}{-210}{280}
\end{center}

\caption{Spectrum of \ser\ at $z = 6.545$, obtained with DEIMOS on the
Keck~II telescope.  The total exposure time is 1~hr, and the spectrum
was extracted using a 1\farcs0 $\times$ 1\farcs2 aperture.  The vertical
dotted line indicates the assumed observed wavelength of \lya.  The dashed
line illustrates the error spectrum, assuming Poisson statistics from
sky plus object.}

\label{fig.1dspec}
\end{figure}

We identified \ser\ in a slitlet targeting CXOSEXSI~J162256.7+264103,
an $f_{\rm 2-10\, keV} = 1.85 \times 10^{-14} \ergcm2s$ hard X-ray
source hosted by an $R \sim 21.7$ galaxy.  Fig.~\ref{fig.image}
illustrates the orientation and size of the slitlet.  Approximately
4\farcs5 from the bright target, an asymmetric, isolated emission line
at 9172 \AA\ is evident in the spectrum.  Fig.~\ref{fig.2dspec} shows
the processed, sky-subtracted, two-dimensional spectrum and
Fig.~\ref{fig.1dspec} presents the extracted, fluxed spectrum of \ser.
We note that since \ser\ is 11\farcm4 from the core of MS1621.5+2640,
gravitational lensing from the galaxy cluster is negligible.  The
extracted spectrum has been corrected for the Galactic extinction of
$E(B-V) = 0.039$, determined from the dust maps of
\markcite{Schlegel:98}Schlegel, Finkbeiner, \& Davis (1998).

\section{Results and Discussion}

\subsection{Spectroscopic Results and Inferred Physical Properties}

\ser\ shows an a single, strong, asymmetric emission line.  In
principle, such high-equivalent-width emission lines may be associated
with several species:  \eg \lya, \oii, \oiii, or H$\alpha$.  In the
case of \ser,  the latter two interpretations can be ruled out by the
lack of neighboring emission lines (H$\beta$ and \oiiia\ for \oiii;
\nii\ and \sii\ for H$\alpha$), while the latter three can be excluded
on the basis of the asymmetric line profile.  In particular,
\oii\ would have been resolved at the resolution of DEIMOS.  Following
the definitions of \markcite{Rhoads:03}Rhoads {et~al.} (2003), we find asymmetry parameters of
$a_\lambda = 3.7$ and $a_f = 2.5$ for \ser, values which are typical of
high-redshift \lya\ emission and are strongly atypical of low-redshift
\oii\ emission \markcite{Dawson:04}(\cf Fig.~3 of Dawson {et~al.} 2004).  Therefore, we
assert that the \ser\ emission line at 9172 \AA\ is unambiguously
identified with \lya, implying a redshift of $z = 6.545 \pm 0.001$,
where the redshift has been measured from the break in the emission
line and the quoted error solely reflects uncertainty in the wavelength
calibration; systematic errors on the redshift due to kinematics and
absorption are likely larger.  Such asymmetric lines are characteristic
of high-redshift \lya\ emission, where outflowing winds cause a broad
red wing, while foreground neutral hydrogen absorption causes a sharp
blue cutoff \markcite{Stern:99e, Dawson:02, Hu:04}(\eg Stern \& Spinrad 1999; Dawson {et~al.} 2002; Hu {et~al.} 2004).

The spectrum of the slitlet target, CXOSEXSI~J162256.7+264103, shows a
composite galaxy spectrum at $z = 0.689$ with several absorption line
features characteristic of early-type galaxies (\eg CaHK, a strong 4000
\AA\ break, G-band) as well narrow emission from \oii\ and \oiiipair.
The emission line of \ser\ would correspond to a rest-frame wavelength
of 5430 \AA\ if it were associated with this galaxy, a wavelength which
does not correspond to any strong features in typical galaxy spectra.
We conclude that the two systems are unrelated.  No continuum source is
evident at the inferred location of \ser\ in the Keck imaging, implying
a 3$\sigma$ limit of $R_E > 26.8$ in a 1\farcs5 diameter aperture.
Since the $R_E$ band corresponds to a rest-frame wavelength of 880
\AA\ for $z = 6.545$, shortward of the Lyman continuum break, this is
consistent with the high-redshift interpretation of \ser.  Furthermore,
no optical sources are evident in our deep ESI image near ($< 4\arcsec$)
from the location of \ser, suggesting that the observed emission line
is unlikely associated with extended \oii\ or \oiii\ emission from
a foreground system.  Such situations are regularly the culprits of
serendipitously-identified, isolated emission lines with more symmetric
profiles \markcite{Stern:00d}(\eg Spinrad, private communication; Stern {et~al.} 2000a).
Finally, we note that since \ser\ lacks an identification in imaging, we
cannot be certain of its location relative to the discovery slitlet and
thus the measured flux suffers some uncertainty from unknown slit losses.

Since the \ser\ \lya\ emission line is asymmetric, we calculate the
line flux by summing the pixels in the range $9170 - 9190$~\AA, finding
a \lya\ flux of $2.13 \pm 0.15 \times 10^{-17} \ergcm2s$.  For $z =
6.545$ and our adopted cosmology, this corresponds to a line luminosity
of $L_{\rm Ly\alpha} = 10.39 \pm 0.73 \times 10^{42} \ergs$.
Determined directly from the pixel flux values, we derive an
observed-frame line FWHM of 6.6 \AA.  Correcting for the instrument
resolution and assuming \ser\ filled the 1\farcs0 wide slitlet, this
corresponds to $\Delta v = 180 \kms$.  For \lya\ emission powered by
star formation, \markcite{Hu:99}Hu, McMahon, \& Cowie (1999) give a conversion rate of $1\, {\rm
M}_\odot\, {\rm yr}^{-1} = 10^{42}\, \ergs$ to relate
\lya\ luminosities to star formation rates \markcite{Rhoads:03, Loeb:04}(but see caveats
in Rhoads {et~al.} 2003; Loeb, Barkana, \& Hernquist 2004), implying a star formation rate of $\geq 10\,
{\rm M}_\odot\, {\rm yr}^{-1}$ for \ser.  The inequality derives from
the unknown fraction of the line flux which has been absorbed by
foreground and associated neutral hydrogen.  We note that studies of
\lya\ emitting galaxies of similar luminosity but at lower redshift
find no evidence of an AGN contribution to the \lya\ luminosity.  In
particular, \markcite{Dawson:04}Dawson {et~al.} (2004) find no \civ\ emission in a composite
spectrum derived from 11 \lya\ emitters at $z \approx 4.5$, while deep
($\sim 170$~ks) exposures with {\it Chandra} of narrow-band selected $z
\approx 4.5$ candidates in the Large-Area Lyman-Alpha (LALA) Bo\"otes
field \markcite{Malhotra:03}(Malhotra {et~al.} 2003) and Cetus field \markcite{Wang:04}(Wang {et~al.} 2004) detect no
X-ray emission, even in stacked X-ray images.

We detect no continuum from \ser.  Averaging the flux in the $9200 -
9300$~\AA\ range, which is relatively clear of telluric emission, we
measure a continuum flux of $-0.03 \pm 0.04 \times 10^{-18}$
\ergAcm2s.  The implied equivalent width of \ser, based on a Monte
Carlo simulation subject to the constraint that both the line flux and
continuum level are positive, is $W_\lambda^{\rm obs} > 410$~\AA\ at
the 67\%\ confidence limit, corresponding to a rest-frame equivalent
width  $W_\lambda^{\rm rest} > 54$ \AA.  For comparison,
\markcite{Dawson:04}Dawson {et~al.} (2004) find a median $W_\lambda^{\rm rest} \approx 90$
\AA\ from a sample of 17 confirmed \lya-emitting galaxies at $z \approx
4.5$.

\subsection{The Surface Density of $z \approx 6.5$ \lya\ Emitters }

The useful sky areal coverage from our three-night DEIMOS observing run
was approximately 5.2 arcmin$^2$, calculated from 1\farcs0 wide slits
covering most of the 16.3\arcmin\ available on each of 19 slitmasks
observed.  Though in principle we are sensitive to emission lines over
a very broad wavelength range \markcite{Cuby:03}(\eg \lya\ emitter at $z = 6.17$
identified by Cuby {et~al.} 2003), we are most sensitive in the
atmospheric-line-free range $9050 \simlt \lambda \simlt 9310$~\AA,
corresponding to \lya\ at $6.44 < z < 6.66$.  The implied surface of
density of \lya\ emitters at $z \approx 6.5$, based on this single
source, is $7.5 \times 10^{-4}$ arcmin$^{-2}$ \AA$^{-1}$, about five
times the surface density derived by \markcite{Kodaira:03}Kodaira {et~al.} (2003) from their
deep, narrow-band imaging survey.  The comoving volume of our survey is
approximately 2600 Mpc$^3$, implying a star formation density of $4
\times 10^{-3}\, {\rm M}_\odot\, {\rm yr}^{-1}\, {\rm Mpc}^{-3}$ at $z
\approx 6.5$.  Since no attempt to account for sources below our
detection threshhold has been made, this might be considered a lower
limit, though we caution against parameters derived from a single
source identified in an apparently unusually fortuitous observation.
For comparison, our estimated star formation density at $z \approx 6.5$
is approximately an order of magnitude greater than that derived by
\markcite{Kodaira:03}Kodaira {et~al.} (2003) and \markcite{Kurk:04}Kurk {et~al.} (2004).  Table~\ref{table} summarizes
the results of several recent surveys for high-redshift
\lya\ emission.

\subsection{Implications for Reionization}

Although it has now been shown that the existence of a {\em single}
\lya-emitting galaxy at a high redshift does not imply that
reionization must have occurred at a yet higher redshift, we do expect
that the epoch of reionization should be accompanied by a rapid decline
in the observed space density of faint \lya\ emitters
\markcite{Haiman:99, Rhoads:01c, Haiman:02}(\eg Haiman \& Spaans 1999; Rhoads \& Malhotra 2001; Haiman 2002).  This is simply due to
the time it takes a young protogalaxy to create and maintain a
sufficiently large \ion{H}{2}, or Str\"omgren, sphere to allow Lyman
photons to escape.  We now consider the luminosity functions of \lya\,
emitters at high redshift.  For simplicity, we omit \lya\ surveys
behind galaxy clusters, where the effects of gravitational lensing
complicate the derived surface densities.

% {\boldmath$z \approx 4.5:$}  Out of a sample of $\sim 350$ candidate
% \lya-emitting galaxies at $z \approx 4.5$ in a search volume of $1.5
% \times 10^6$ co-moving Mpc$^3$ (\ie 2 fields, $36\arcmin\, \times\,
% 36\arcmin$ each, sampling \lya\ emission at $4.37 < z < 4.57$),
% \markcite{Dawson:04}Dawson {et~al.} (2004) present spectroscopic results for 25 sources, out
% which 18 are confirmed to reside at the target redshift. 

{\boldmath$z \approx 5.7:$}  There are two recent reports on
spectroscopic confirmation of $z \approx 5.7$ \lya-emitting galaxies.
\markcite{Rhoads:03}Rhoads {et~al.} (2003) report on 18 candidate $5.67 < z < 5.80$
\lya\ emitters in 710 arcmin$^2$ survey region, out of which 4 sources
were attempted spectroscopically and 3 were confirmed.  \markcite{Hu:04}Hu {et~al.} (2004)
present results on 26 narrow-band selected candidate $5.653 < z <
5.752$ \lya\ emitters in a 702 arcmin$^2$ region, out of which 23
sources were attempted spectroscopically and 19 were confirmed.
\markcite{Hu:04}Hu {et~al.} (2004) do not provide the spectroscopically-derived
\lya\ fluxes, instead giving the fluxes of the sources in a narrow-band
filter ($f_\nu^{\rm NB}$; $\lambda_c = 8150$\AA, $\Delta \lambda_{\rm
FWHM} = 120$\AA) and in a long-pass $z$ filter ($f_\nu^{\rm z}$).  We
model the narrow-band filter as a top-hat function, the \lya-emitter
galaxy spectra as a \lya\ emission line of flux $f_{\rm Ly\alpha}$
superimposed upon a flat (in $f_\nu$) continuum longward of \lya, and
we assume an opaque \lya\ forest at this redshift implying negligible
flux blueward of \lya.  In this approximation, the \lya\ flux of a
source at redshift $z$ is given by $$ f_{\rm Ly\alpha} = {{c} \over
{\lambda_{\rm Ly\alpha}}^2} \left[ 120 {\rm \AA} f_\nu^{\rm NB} - (8210
{\rm \AA} - {\lambda_{\rm Ly\alpha}}) f_\nu^{\rm z} \right], $$ where
${\lambda_{\rm Ly\alpha}} \equiv 1216\, (1 + z)$ \AA.

{\boldmath$z \approx 6.5:$}  Several \lya\ emitters at this redshift have
been reported in the literature from a variety of different surveys.
We attempt to combine these disparate results into a single cumulative
luminosity function.  \markcite{Kodaira:03}Kodaira {et~al.} (2003) report on 73 candidate $6.508
< z < 6.617$ \lya\ emitters in an 814 arcmin$^2$ survey region, out of
which 9 sources were attempted spectroscopically and 2 were confirmed.
For our adopted cosmology, this survey samples a co-moving volume
of $2.02 \times 10^5$ Mpc$^3$, though correcting for spectroscopic
incompleteness, the effective volume is 9/73 of this.  \markcite{Kurk:04}Kurk {et~al.} (2004)
report on a single source at $z = 6.518$ in a slitless survey over 43
arcmin$^2$ which was sensitive to \lya\ emission at $6.4 < z < 6.6$.
\markcite{Rhoads:04}Rhoads {et~al.} (2004) report on four candidate $6.52 < z < 6.55$ \lya\
emitters in a 1200 arcmin$^2$ survey region.  All four sources have been
followed up spectroscopically, and only one is confirmed at high redshift.
Finally, the relevant parameters for the survey which identified \ser\
are provided in \S~3.2.

\begin{figure}[!t]
\begin{center}
\plotfiddle{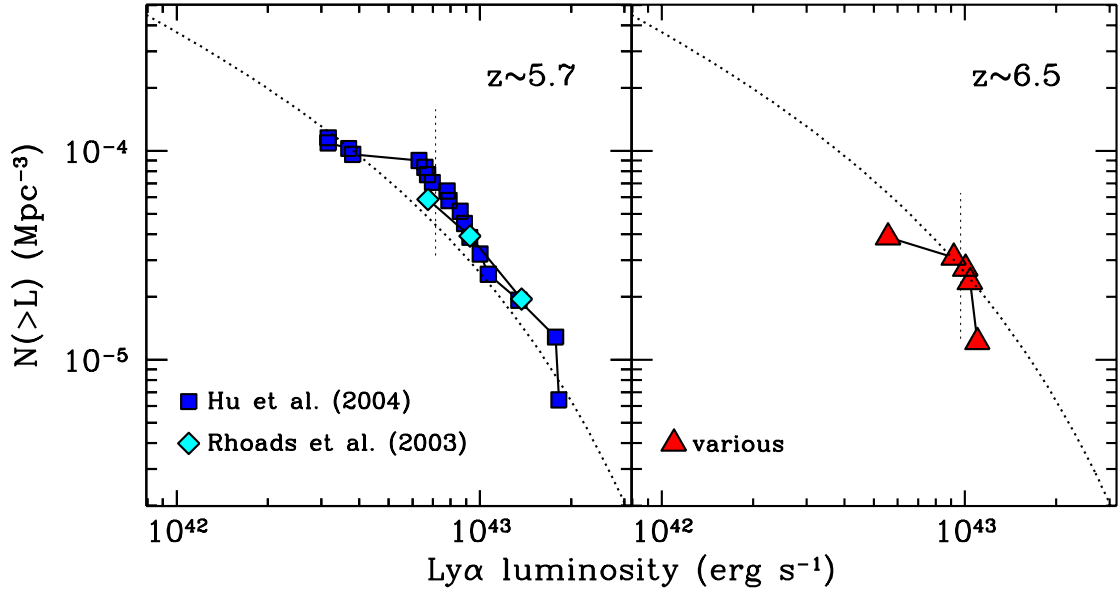}{2.6in}{-90}{60}{60}{-220}{300}
\end{center}

\caption{The empirical cumulative luminosity functions for
\lya\ emitters at $z \sim 4.5$, $z \sim 5.7$, and $z \sim 6.5$.  All
points are from surveys with spectroscopic follow-up.  The dashed curve
illustrates a non-evolving Schechter luminosity function with $L^* =
1.4 \times 10^{43} \ergs$, $\alpha = -1.6$, and $\phi^* = 1.0 \times
10^{-4}$ Mpc$^{-3}$, and is meant as only a fiducial line for comparing
the panels.  The vertical dotted lines illustrate the line luminosity
corresponding to a line flux of $2 \times 10^{-17} \ergcm2s$ for each
redshift range.  This corresponds to the rough limit of most of the
surveys.}

\label{fig.lf}
\end{figure}

Fig.~\ref{fig.lf} presents the empirical cumulative luminosity function
of \lya\ emitters in these two redshift windows, along with a fiducial
non-evolving Schechter luminosity function intended for comparison
purposes.  Only minimal attempts to account for incompleteness have been
attempted here; we therefore expect large errors both at the bright end,
from small number statistics, and at the faint end, from incompleteness.
Restricting the analysis, therefore, to moderate luminosity sources, we
find no strong evolution in the cumulative luminosity function of \lya\
emitters between $z \approx 5.7$ and $z \approx 6.5$.  A more detailed
analysis of the luminosity function of \lya\ emitters in these two
redshift windows is provided by \markcite{Malhotra:04}Malhotra \& Rhoads (2004); they also find a
lack of evolution.  If the reionization epoch represented a very large
{\it and} rapid change in the neutral fraction of the IGM, this event
should produce an associated change in the observable properties of \lya\
emitters \markcite{Gnedin:00}(\eg Gnedin 2000).  For example, in the dynamic IGM models
of \markcite{Santos:04}Santos (2004), as $x_{\rm HI}^{\rm IGM}$ increases from 0.05 to
0.95, the observed \lya\ line flux drops by nearly an order of magnitude.
In this picture, the high surface density of \lya\ emitters at $z \approx
6.5$ therefore suggests that the IGM does not evolve rapidly between $z
\approx 5.7$ and $z \approx 6.5$.  However, if the reionization transition
was more gradual, depending on the luminosity and density evolution of
ionizing sources, or if there were some geometric complexities, \eg,
due to a bias-driven clustering of sources, the observable changes in
the apparent luminosity function of Ly$\alpha$ emitters could have been
more gradual.

\section{Conclusions}

We report the discovery of a \lya-emitting galaxy at $z = 6.545$
serendipitously identified with the Keck~II telescope.  The galaxy
resides within 30\arcsec\ of a bright ($R = 12$) Galactic star, and may
provide a useful target for adaptive optics imaging.  Our discovery
marks the sixth \lya-emitting galaxy identified at $z \approx 6.5$.
This redshift lies beyond the putative epoch of reionization as
inferred from the spectra of high-redshift quasars, when the damping
wings of the neutral IGM should severely attenuate \lya\ emission.
Models predict that \lya\ emission from individual sources of
sufficient age and luminosity to create \ion{H}{2} regions in a neutral
IGM might still be visible beyond the epoch of reionization.  However,
the expectation is that the luminosity function of \lya\ emitters, at
least at the faint end, should drastically change as this threshold is
crossed.

We combine results from the literature to compare the cumulative
luminosity function of \lya\ emitters at $z \approx 5.7$ and $z \approx
6.5$, finding no evidence of dramatic change, at least in the luminosity
range probed here.  Similar conclusions are reached in the more detailed
analysis of \markcite{Malhotra:04}Malhotra \& Rhoads (2004), a coordinated paper which derives the
luminosity functions in these two redshift windows.  These results
suggest that the universe remains largely ionized to $z \approx 6.5$.

This does not necessarily contradict the quasar results.  The observed
Gunn-Peterson troughs can be caused by partly reionized regions of the
IGM with neutral hydrogen fractions as low as $x_{\rm HI}^{\rm IGM}
\sim 10^{-3}$, whereas the counts of \lya\ sources can probably easily
accommodate $x_{\rm HI}^{\rm IGM} \sim 10^{-1}$, which may be also
consistent with the results by \markcite{Wyithe:04}Wyithe \& Loeb (2004a) and
\markcite{Mesinger:04}Mesinger \& Haiman (2004).  Also, whereas some observers find a qualitative
change in the absorption properties of the IGM at $z \sim 6$
\markcite{Fan:02, Cen:02, White:03}(\eg Fan {et~al.} 2002; Cen \& McDonald 2002; White {et~al.} 2003), others do not
\markcite{Songaila:04}(Songaila 2004).  The nature, extent, and geometry of the
reionization transition from $x_{\rm HI}^{\rm IGM} = 1$ to $x_{\rm
HI}^{\rm IGM} < 10^{-3}$ remain highly uncertain.

We caution that these results on the luminosity function of \lya\
emitters at $z \approx 6.5$ rely somewhat on the serendipitous
discovery of \ser\ described herein.  As the surface density of $z
\approx 6.5$ \lya\ emitters inferred by \ser\ is somewhat (a factor of
five) larger than that derived from other surveys, the results
described here might instead be interpreted to imply that at least a
subset of the authors of this manuscript are lucky, a result that has
been hinted at in previous work \markcite{Becker:92, Dey:98,
Stern:00c, Dawson:02, Stern:04a}(\eg Becker, Helfand, \& White 1992; Dey {et~al.} 1998; Stern {et~al.} 2000b; Dawson {et~al.} 2002; Stern {et~al.} 2004).

We emphasize that the surface density of \lya\ emitters offers an
independent tool for studying the ionization state of the IGM,
sensitive to ionization fractions $0.1 \simlt x_{\rm HI}^{\rm IGM}
\simlt 1$ expected near the epoch of reionization, ionization fractions
which are difficult to probe with quasar absorption studies.  As more
\lya\ emitters are confirmed at high redshift, the statistical weight
of the results hinted at here will be tested.  We note, however, that
models of protogalaxies suggest winds might compromise the ability of
\lya\ emission to constrain strongly the neutral fraction of the IGM
without restframe optical observations to constrain the systemic
redshift, age, and luminosity of the young galaxies \markcite{Santos:04}(Santos 2004).
Furthermore, biased clustering around \lya\ emitters likely contributes
significantly to the ionizing photon budget for the more massive
protogalaxies \markcite{Wyithe:04b}(\eg Wyithe \& Loeb 2004b); changes in the \lya\ emitter
luminosity function prior to the epoch of reionization are potentially
only significant at the lowest luminosities.  Speculatively, we
conclude by noting that only one source has been identified 
beyond $z \approx 6.5$ to date, a lensed source at $z \approx 7$ found
by \markcite{Kneib:04}Kneib {et~al.} (2004)\footnote{The lensed $z = 10.0$ source identified by
\markcite{Pello:04}Pell\'o {et~al.} (2004) has recently come into question based on a reanalysis
of the discovery spectroscopic data by \markcite{Weatherley:04}Weatherley, Warren, \&  Babbedge (2004) and deep
near-infrared imaging by \markcite{Bremer:04}Bremer {et~al.} (2004) which fails to detect the
source.}.  The \markcite{Kneib:04}Kneib {et~al.} (2004) source lacks \lya\ emission, making the
exact redshift identification extremely difficult.  This is suggestive
of a source identified beyond the epoch of reionization, and may be the
first example of the challenges of observational astronomy as we
attempt to identify and study sources embedded in a neutral Universe.

\acknowledgements 
We gratefully acknowledge enlightening comments provided by A.~Loeb,
B.~Becker, Z.~Haiman, L.~Hui, and H.~Spinrad on earlier drafts of this
manuscript, as well as the insights provided by the shared history of
studying similar sources with S.~Dawson, A.~Dey, B.~Jannuzi, and
H.~Spinrad.  The authors wish to recognize and acknowledge the very
significant cultural role and reverence that the summit of Mauna Kea
has always had within the indigenous Hawaiian community; we are most
fortunate to have the opportunity to conduct observations from this
mountain.  The work of DS was carried out at Jet Propulsion Laboratory,
California Institute of Technology, under a contract with NASA.

%% TABLE 1
\begin{deluxetable}{ccccccl}
%\tablewidth{0pt}
\tablecaption{Surface density of \lya\ emitters}
\tablehead{
\colhead{} &
\colhead{Survey Area} &
\colhead{Flux Limit} &
\colhead{\#} &
\colhead{Spec.} &
\colhead{$\Sigma({\rm Ly}\alpha)$} &
\colhead{} \nl
\colhead{$z$} &
\colhead{(arcmin$^2 \times$\AA)} &
\colhead{(\ergcm2s)} &
\colhead{Cand.} &
\colhead{Success} &
\colhead{(${\rm deg}^{-2}\, {\rm unit-}z^{-1}$)} &
\colhead{Reference}}
\startdata
3.4 & $  75 \times  77$ & $1.5\times10^{-17}$ &  19 & 15/15 & 14,400 & \markcite{Hu:98}Hu {et~al.} (1998) \\
\\
4.5 & $  25 \times  78$ & $1.5\times10^{-17}$ &   3 &  2/3  &   4490 & \markcite{Hu:98}Hu {et~al.} (1998) \\
    & $1116 \times  85$ & $2.6\times10^{-17}$ & 225 &  1/3  &   3460 & \markcite{Rhoads:00}Rhoads {et~al.} (2000) \\
%     & $2592 \times 243$ & $\approx 2\times10^{-17}$ &$\sim 350$ & 18/25 & $\sim 1750$ & \markcite{Dawson:04}Dawson {et~al.} (2004) \\
\\
5.7 & $ 710 \times 150$ & $\approx 2\times10^{-17}$ &  18 &  3/4  &    560 & \markcite{Rhoads:03}Rhoads {et~al.} (2003) \\
    & $ 702 \times 120$ & $2\times10^{-17}$ &  26 & 19/23 &   1120 & \markcite{Hu:04}Hu {et~al.} (2004) \\
    & $   5 \times 150$ & $6\times10^{-18}$ &   0 &  0/0  & $<5840$& \markcite{Martin:04}Martin \& Sawicki (2004) \\
\\
6.5 & $ 814 \times 132$ & $\approx 1\times10^{-17}$ &  73 &  2/9  &    660 & \markcite{Kodaira:03}Kodaira {et~al.} (2003) \\
    & $1200 \times  84$ & $2\times10^{-17}$ &   4 &  1/4  &     40 & \markcite{Rhoads:04}Rhoads {et~al.} (2004) \\
    & $  18 \times 190$ & $5\times10^{-18}$ &   0 &  0/0  & $<1280$& \markcite{Tran:04}Tran {et~al.} (2004) \\
    & $  43 \times 190$ & $\approx 2\times10^{-17}$ &   1 &  1/1  &    540 & \markcite{Kurk:04}Kurk {et~al.} (2004) \\
    & $   5 \times 260$ & $\approx 2\times10^{-17}$ &   1 &  1/1  &   3370 & this paper \\
\enddata

\tablecomments{We omit \lya\ surveys behind galaxy clusters, where the
effects of gravitational lensing complicate the derived surface
densities.  We also omit imaging surveys which have not yet reported on
spectroscopic success rates.}

\label{table}
\end{deluxetable}

%% FIGURE 1

%% FIGURE 2

%% FIGURE 3

%% FIGURE 4

%\bibliographystyle{apj.bst}
%% \bibliography

% FIGURES

\clearpage
\end{document}